# Segmentation of Lungs in Chest X-ray Image using Generative Adversarial Networks

FAIZAN MUNAWAR[1], SHOAIB AZMAT[1], TALHA IQBAL[2], CHRISTER GRONLUND[3], HAZRAT ALI[1*]

[1] Department of Electrical and Computer Engineering, COMSATS University Islamabad, Abbottabad Campus, Pakistan.
[2] Department of Medicine, National University of Ireland, Galway, Ireland.
[3] Department of Radiation Sciences, Biomedical Engineering, Umeå University, Umeå, Sweden.
faizan.munawar007@gmail.com, shoaibazmat@cuiatd.edu.pk, t.iqbal1@nuigalway.ie, christer.gronlund@umu.se, hazratali@cuiatd.edu.pk

Corresponding author: Hazrat Ali: hazratali@cuiatd.edu.pk

**ABSTRACT** Chest X-ray (CXR) is a low-cost medical imaging technique. It is a common procedure for the identification of many respiratory diseases compared to MRI, CT, and PET scans. This paper presents the use of generative adversarial networks (GAN) to perform the task of lung segmentation on a given CXR. GANs are popular to generate realistic data by learning the mapping from one domain to another. In our work, the generator of the GAN is trained to generate a segmented mask of a given input CXR. The discriminator distinguishes between a ground truth and the generated mask, and updates the generator through the adversarial loss measure. The objective is to generate masks for the input CXR, which are as realistic as possible compared to the ground truth masks. The model is trained and evaluated using four different discriminators referred to as D1, D2, D3 and D4 respectively. Experimental results on three different CXR datasets reveal that the proposed model is able to achieve a dice-score of 0.9740, and IOU score of 0.943, which are better than other reported state-of-the art results.

## I. INTRODUCTION

With ever increasing capabilities of modern technologies, the role of computer-aided diagnosis (CAD) systems has reached to a maximum significance than ever before [51]. Today, CAD systems provide aid to physicians and health care professionals to better understand the clinical conditions of a patient and help in diagnosis of various diseases. Medical image procedures such as magnetic resonance imaging (MRI), X-rays and ultrasound imaging carry useful information, but require thorough study by an expert. Typically, a radiologist would need to manually examine and extract the useful information from an x-ray while adhering to time constraints, for example in overly crowded hospitals (as most common in developing countries), or in case of a pandemic (as the recent surge in covid-19 cases proving the same). Automatic CAD with machine learning helps in speeding up the process and easing out the workload on health professionals. Typically, CAD systems are used in various areas of medical diagnosis such as in mammography, for detection of breast cancer, polyps detection in colon, diabetic retinopathy, lung cancer detection, coronary artery disease detection and pathological brain detection. Computer programs have been developed to help in the detection of various diseases [52].

CAD system can be divided into three categories. First is classification, second is detection and the last one is segmentation. In classification, the objective of the system is to classify each image in any of the pre-defined categories, such as an image with a tumor and an image with no tumor. In detection, the system is programmed to detect a certain localized region, such as detection of tumors in brain. The resulting image typically contains a bounding box to highlight the tumor. The segmentation goes one step further and divides the image into different categories on a pixel level. Each pixel represents its respective category, such as pixels of tumor and pixels of the region with no tumor. Segmentation lies at the heart of automatic CAD systems. Multiple imaging techniques are used for creating medical images including X-rays, MRI scans, ultrasound, CT (computed tomography) scans and PET (positron emission tomography) scans. Chest X-ray (CXR) procedure is often two to ten times more common than other medical imaging procedures such as MRI, CT scans and PET scans [1]. Typically, several thousand CXRs are generated in a large hospital, significantly raising diagnostic workloads. Furthermore, detecting the lung region in CXR images is an important component in a CAD of lung health. Figure 1 shows example CXRs and the corresponding segmented lung region.

One of the most important steps in automatic analysis of CXRs is to detect the lung boundaries accurately. The boundary extraction is the key to identify shape irregularity and lung volume, and thus provides insight into identification of cardiomegaly, pneumothorax, pneumoconiosis or emphysema [2]. The strong edges appearing at the rib cage and clavicle region, variations in heart anatomy, and non-homogeneities in the imaging make the segmentation of lungs challenging. These factors usually push the minimization approach to end up at local minima [3].

Segmenting lungs also aids in reducing computational complexity for other lung related disease identification algorithms by performing computations only on lung regions. Additionally, segmenting lungs is very critical in lung related disease identification such as pneumonia [4] and Tuberculosis [5], [6]. Many authors have demonstrated the importance of lung segmentation prior to further processing e.g. authors in [7] segmented the lung regions before applying further processing for Covid19 detection. In another work, the authors in [8] have used CT slices of lungs to detect Covid19 infection using a series of processing techniques [8]. Authors in [9] proposed a technique for circular object detection in CXR's and demonstrated that lung segmentation increases the overall performance. In [10] authors proposed a model for detecting rotation in CXR's and reported that when lungs in CXR are poorly segmented, the accuracy of the model drops rapidly. Authors in [11] proposed a technique for detection of pulmonary abnormalities and showed that lung segmentation is a crucial step before any further classification.

The segmentation of lungs from CXR images falls into two general categories: (1) Conventional methods (2) Deep learning based methods. Conventional methods can be categorized in to four categories: (i) Rule based methods [1], (ii) Pixel classifier-based methods [12], [13], (iii) Deformable models, and (iv) Hybrid approach [12], [14], [15]. In deep learning based methods, there are two major categories: (i) Discriminative models e.g. U-net [16], and (ii) Generative models e.g. auto encoders, generative adversarial networks [17] [18] [19].

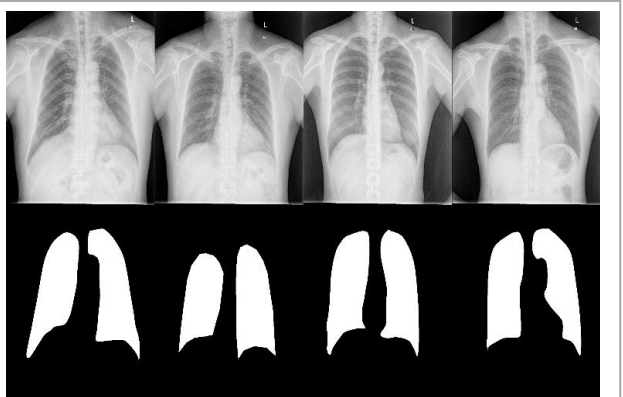

FIGURE 1. X-ray images and corresponding segmented lung masks

First, discussion on conventional methods is presented. Rule based models contain simple techniques such as thresholding and morphological operations. In [20], the authors used rule based anatomical information for lung region extraction.

Pixel classification models work by classifying pixel as either inside or outside the lung region. The authors in [12] used a contrast difference between lung fields and borders to orientate an active contour technique.

Deformable models are very popular and extensively studied. They are applied extensively in medical image segmentation, [21]. Authors in [22] proposed an optimized "active shape model" (ASM) for lung segmentation. They addressed the problem of typical ASM, which was the need of sufficient initialization close to the target. They reported significant improvement in accuracy compared to a typical ASM.

Hybrid models are composed of multiple conventional models. The authors in [23] proposed a robust hybrid method for lung segmentation. They combined a rule-based method with pixel classification approach. Using these two techniques on a dataset of 230 images, an accuracy of 94% is reported in [23].

Next, deep learning based methods are presented. In [25] authors proposed a deep learning based method for the task of semantic segmentation. They used existing AlexNet and VGG-Net models for classification, customized them to be fully convolutional networks, and then performed segmentation by using the output of these networks. In [16], the authors proposed a network for the task of segmentation of neuronal structures in electron microscopic stacks. They named their network U-net which is a ladder like structure made up of two parts, encoder and decoder. They also used skip connections between encoder and decoder layers that helped in improving the performance of overall segmentation. In [26], the authors proposed a deep learning based lung segmentation approach. They name their network as SegNet which is a fully connected deep neural network. They used encoder-decoder approach designed in such a way that low resolution features are mapped to input resolution for pixel-wise classification, producing features that aid in accurate boundary localization. In [27], the authors proposed a modified U-Net network using residual structure for segmentation of Covid-19 CT images. Use of residual structure improved the feature selection process and thus improved the overall segmentation. In [28], the authors proposed COVID-19 lung CT infection segmentation network, named as Inf-Net that uses reverse attention and explicit edge attention. The authors claimed that their network achieved higher accuracy than the other methods.

In [53], the authors proposed a patch based deep belief networks for vertebrae segmentation in CT images. They have shown that the patch-based model achieved superior performance compared to the traditional techniques.

Along with being popular for image generation, generative adversarial networks make up another set of methods used for segmentation [17], [27], [29-33]. In [54], the authors proposed a variation of generative adversarial network named as hybrid fusion network, for generating synthetic MR image modalities. Due to patients' dropouts and poor quality of data, missing modalities is a huge challenge. The authors have shown that their network has outperformed state of the art synthesis networks by achieving higher accuracy [54]. In [24], the authors proposed an unsupervised adversarial similarity network for image registration. They proved that their network can train without the ground truth, and can achieve state of the art results for image registration on brain MRI images. The authors in [34] trained a GAN framework to perform segmentation of liver images. In [35], the authors trained a GAN to produce lung nodules and then using the synthetic data, trained a progressive holistically nested network (P-HNN), which is a segmentation network. The generator was conditioned on volume of interest and thus, it was able to generate realistic 3D lung nodules. Both the robustness and accuracy of a P-HNN have been reportedly improved. In [36], the authors proposed a framework for the generation of vessel maps from the retinal images. The authors also formulated the objective function of GAN for the segmentation task. They achieved competitive dice score on DRIVE and STARE datasets of retinal fundoscopic images [36]. In MI-GAN [37], the authors proposed a framework for the generation of synthetic medical images as well as their segmented masks. The synthetic images and their masks are further used for training of segmentation network, and the authors reported state-of-the-art dice score on DRIVE and STARE datasets [37]. In [38], the authors showed that augmenting data via GANs can improve the overall segmentation result, especially when there is limited data for the training. They performed experiments on MR and CT images of brain [38]. In [39], the authors proposed a model for generating CT images from input MRI images. They addressed the issue of a blurry CT when generated, and thus proposed a loss function that deals with this issue of generated CT images. In [40], the authors demonstrated the generation of CXR and their segmented masks. They showed that when the pair is generated, the quality of images is reduced as compared to the case if only a CXR image is generated [40].

Lung segmentation task can be considered as an image translation problem, where segmented masks are produced at output using CXR image as an input to the model, because the domain of the input image is changed from a gray scale X-ray image to a binary mask image, as reported in [42]. However, GANs have not been developed for lung image segmentation.

In this work, GAN based model for segmentation of lungs from Chest X-ray images is presented. Our model includes a generator similar to U-net architecture, [37] and multiple discriminators which are binary classifiers, which has not been reported to the best of our knowledge.

Our main contributions are:
- We present a GAN based approach for segmentation of lungs in chest x-ray images.
- We train the generator in a way such that the generator learns the distribution $G_\theta$ from very small training dataset.
- We report the segmentation results on different types of discriminator networks to identify the best possible match and achieved better accuracy than the state of the art. To the best of our knowledge, there has been no work reported for lung segmentation with U-net like generator and comparative analysis based on multiple discriminators.

In this work, 97.4% mean dice score on lung segmentation task has been achieved, which is better than the state-of-the-art dice score.

The rest of the paper is organized as follows. In Section II, generative adversarial networks are discussed. Additionally, the proposed architecture for segmentation is discussed in the same section. In Section III, experimental setup and data description are provided. Section IV presents the results and discussion. Finally, the paper is concluded in Section V.

## II. METHODOLOGY

In this section, overall architecture for the GAN is discussed and an explanation on the general working of GAN as well as the specific GAN model and the underlying discriminators used are presented.

### A. GENERATIVE ADVERSARIAL NETWORKS

Goodfellow first proposed an adversarial process for generative models named as the GAN [17]. A GAN model consists of Generator G and a discriminator D. The input to G is random noise vector and G tries to generate data which fits the real data distribution. The function of D is to authenticate the data i.e., whether it is real or fake. Experiments were performed on different datasets including MNIST and Toronto face dataset (TFD). Different variants of GANs were proposed after that, in which most of them tried to stabilize GAN training [29-33].

In [41], the authors introduced deep convolutional GAN (DCGAN), which proposed certain constraints on the architecture topology to make GANs stable. They demonstrated high accuracy of discriminator for classification, and also demonstrated the vector arithmetic properties of generators [41]. The work published in [42] proposed image to image translation, a GAN framework in which the generator is conditioned, i.e., a sample distribution along with the random noise is also input to the generator. The GAN then learns the mapping from input image to output image, thus generating images of certain desirable distribution [42]. Generative Adversarial Network (GAN) as first proposed by Goodfellow [17], is a combination of two neural networks trained in an adversarial fashion. A GAN consists of a generator G network and a discriminator D network. The purpose of the generator is to generate data from random noise, and the discriminator discriminates between real data and the data generated by the generator.

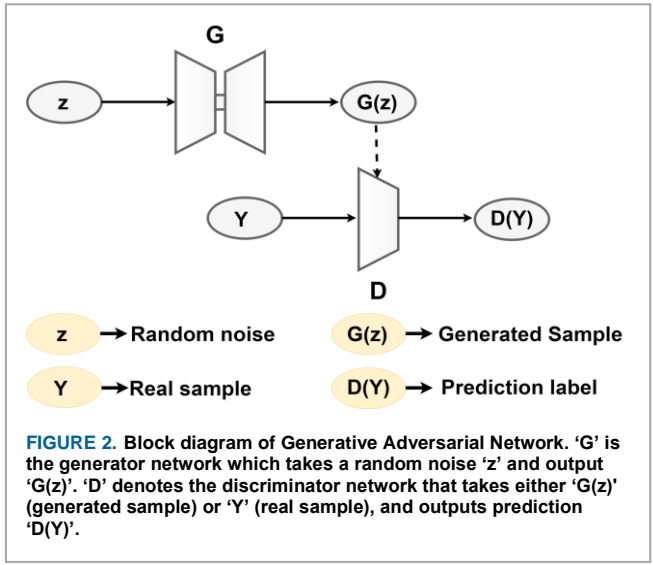

**FIGURE 2.** Block diagram of Generative Adversarial Network. 'G' is the generator network which takes a random noise 'z' and output 'G(z)'. 'D' denotes the discriminator network that takes either 'G(z)' (generated sample) or 'Y' (real sample), and outputs prediction 'D(Y)'.

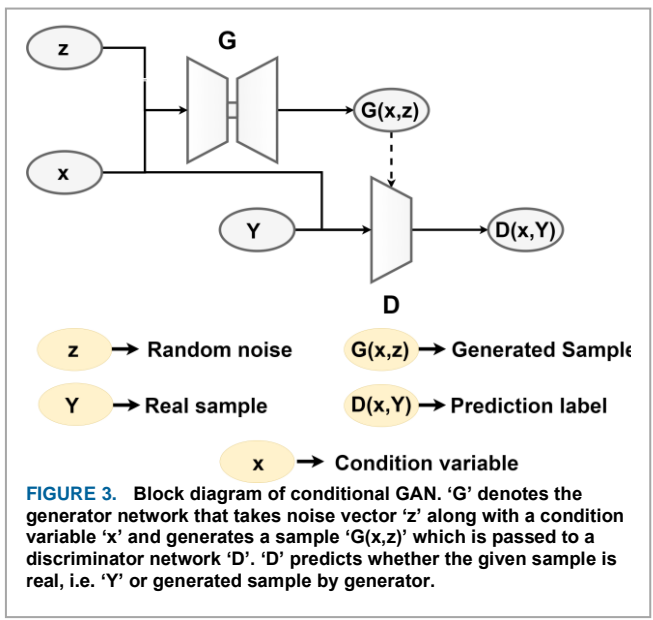

**FIGURE 3.** Block diagram of conditional GAN. 'G' denotes the generator network that takes noise vector 'z' along with a condition variable 'x' and generates a sample 'G(x,z)' which is passed to a discriminator network 'D'. 'D' predicts whether the given sample is real, i.e. 'Y' or generated sample by generator.

The training of GAN can be considered as a dual player min max game. During the training, the generator G tries to generate realistic looking data such that the discriminator D can be fooled. However, the discriminator tries to efficiently classify the real and the fake data. Thus G tries to minimize the possibility of being predicted as fake, and D tries to maximize the probability of correct predictions. The general structure of GAN is shown in Figure 2.

The objective function of original GAN as described in [17] is given by minimization and maximation of *V*, as represented in Eq 1 (as in [17]):

$$\min_G \max_D V(G, D) = \mathbb{E}_y[\log D(y)] + \mathbb{E}_z[\log(1 - D(G(z)))] \quad (1)$$

Where, *y* represents the real data, *z* is the random noise vector, *D(y)* is the discriminator's prediction on y, *G(z)* is the generated data from noise *z*, and *D(G(z))* provides the discriminator's prediction on generated data.

Both the generator and discriminators play the min-max game and adjust their parameters according to each other's actions. Over a certain number of training iterations, there is a possibility that both the networks will have parameters which may not be further optimized. At such a stage, the generator generates very realistic looking synthetic data, such that the discriminator is not able to differentiate between the real and generated data.

In order to generate a specific domain data, certain constraints can be introduced on the generator as proposed in conditional GAN (cGAN) model [42]. In cGAN, the training is conditioned on an additional input variable *x*. The optimization problem then becomes (as in [42]):

$$\min_G \max_D V(G, D) = \mathbb{E}_{x,y}[\log D(x,y)] + \mathbb{E}_{x,z}[\log(1 - D(x, G(x,z)))] \quad (2)$$

Where, *x* is the conditioning variable. Structure of conditional GAN is shown in Figure 3.

We use the cGAN model to perform lung segmentation. In our model, the generator is conditioned on the CXR image to produce segmented lungs as shown in Figure 4. The optimization problem then relies on reducing the L1 loss calculation as used by [35], [37]. Our optimization L1 loss is given as:

$$L_{L1loss} = \mathbb{E}_{x,y,z} | y - G(x,z) | \quad (3)$$

L1 loss penalizes the distance between segmented images and their ground truth masks. Optimization of L1 loss ensures that the generated image by generator 'G' is not much deviated from the segmentation mask provided, so our final objective function becomes:

$$\min_G \max_D GAN = V(G, D) + \alpha L_{L1loss} \quad (4)$$

Where α is the trade-off constant with α > 0. Now, an explanation on the proposed generator and discriminator architecture is presented in detail.

### B. GENERATOR ARCHITECTURE

Most works on the GANs have reported the use of an encoder-decoder style structure for generator network. So, a similar architecture for the generator is retained [37][43]. This allows us to use noise code in a natural manner. Encoder extracts features of the input image, as it is a multiple layered neural network. It extracts the local features in the first few layers, and as it goes deeper it extracts the more global information. The first few layers of the encoder down-samples the input until a bottleneck, after which decoder up-sampling takes place.

In the first four layers of our generator G, the input is down-sampled and features are learned. Whereas the fifth layer is the bottleneck layer, after which up sampling starts and the image is reconstructed as shown in Figure 5. As explained in [42], our architecture does not need random noise because it learns to ignore it, and hence does not improve results.

Another important part of the generator is skip connections, as used in U-Net [16]. Skip connections are used for various purposes as explained in the section below.

## 1) SKIP CONNECTIONS

Skip connections of U-Net are used as this helps to preserve the morphology and reduce the risk of vanishing gradient. Skip connections were initially proposed in residual networks [44]. The skip connections allow better generalization by facilitating the direct passing of error gradients between decoder-encoder layers. Next, different types of discriminators used in this work are discussed.

## C. DISCRIMINATOR

Discriminator is a deep convolution neural network that classifies its input image as real or fake. The network architecture is of the form convolution – batch normalization – activation. Input to the discriminator is either the lung mask (image) generated by the generator network, or the lung mask from the ground truth. The discriminator classifies each image as either synthetic or real. Discriminator's configurations as either image discriminator or patch discriminator are utilized [42]. The main difference between image discriminator and patch discriminator is that the image discriminator maps the input image to a single scalar output and thus the prediction is based on the entire image, whereas the patch discriminator divides the input image to a set of patches and maps each patch to a single scalar output.

Prediction is made on each patch of size N×N pixels. For patch discriminators, there is no overlap between patches. For final classification, individual result of each patch is averaged. Patch discriminators effectively model input image as Markov random field, assuming independency between pixels and patches.

As reported by [42], the patch discriminator solves the problem of blurry images caused by failures at high frequencies like edges. Another advantage of a patch discriminator is that it has fewer parameters than the full image discriminator. Thus, it reduces computational time and can work better for arbitrary very large images. The general structure of the discriminator is shown in Figure 6. Details of different discriminators in our work are provided below.

## 1) DISCRIMINATOR D1

This discriminator has the patch size of 1×1 pixels (PixelGAN). The Discriminator consists of one input layer, three hidden layers and a final output layer. For the final layer, sigmoid activation function has been utilized, whereas for the hidden layers, leaky ReLU is used. This discriminator tests every single pixel and classifies it as '1' or '0'. The final prediction is made based upon the average result of each individual pixel of the input image.

## 2) DISCRIMINATOR D2

For the second discriminator, the structure of patchGAN, having a patch size 16×16, is used. The network is of the form conv-batch norm-relu with five hidden layers. The difference between patchGAN and the PixelGAN is that rather than a prediction on individual pixels, it predicts the patch of image of size N×N where N≠1. The final prediction is made based upon the average prediction of all the patches in the input image.

## 3) DISCRIMINATOR D3

For the third discriminator, the patch size is varied from 16×16 to 70×70 pixels. In literature, 70×70 patch size has shown better results because 70×70 patch size is big enough to accommodate global features, however, the number of parameters are reduced because of lesser size than the whole image [42]. The network has three hidden layers and is of the form conv-batchnorm-relu.

## 4) DISCRIMINATOR D4

For the fourth discriminator, we use the full image discriminator. It gives the final prediction based upon the whole image rather than patches. The network consists of one input layer, five hidden layers of the form convolution-batchnormaliztion-relu and a final output layer.

For all the discriminators, stride and kernel size are fixed with values 2 and 3×3, respectively.

## III. EXPERIMENTAL SETUP AND DATA DESCRIPTION

All the experiments were performed using a standard PC with Intel core i7 CPU and GeForce GTX 1080 GPU with 8 GB memory. Python was used as programming language because of its vast usage and built-in easy to use libraries. The python implementation relies on the Keras and Tensorflow libraries and the MI-GAN code base with necessary modifications. For optimization of networks, Adam optimizer was used with a learning rate of 0.0002 and decay rate of 0.5, because they have shown superior performance in our initial experimentation.

For evaluating our networks, the three most commonly used datasets of Chest X-rays were used. The first dataset is the JSRT dataset, the second one is the Montgomery dataset, and the third is the Shenzhen Chest X-ray dataset. Images in these datasets differ in terms of the resolution and contrast. All these datasets contain 2D chest x-ray grayscale images recorded at various hospitals. The JSRT dataset was developed in 1998 in collaboration with the Japanese Radiological Society (JRS) by the "Japanese Society of Radiological Technology" (JSRT) [45]. The X-ray images in the Montgomery dataset were collected by the "Department of Health and Human Services of Montgomery County, MD, USA", as part of the Tuberculosis control program. The third dataset, the Shenzhen was created as a result of a partnership between "Shenzhen No. 3 People's Hospital, Guangdong Medical College", China, the "National Library of Medicine, Maryland, USA" [46]. All three datasets were free from any noise, so no preprocessing was used. Table 1 provides summary of the x-ray images with their specifications. Sample images are shown in Figure 8, 9 and 10.

## IV. RESULTS AND DISCUSSION

For evaluating our segmentation network, two parameters dice score and intersection over union (IOU) are calculated, as these are widely accepted evaluation parameters for segmentation techniques. Dice score is defined as:

$$Dice\ coefficient = \frac{2\ TP}{2TP+FP+FN}$$

and IOU is calculated by:

$$IoU = \frac{TP}{TP+FP+FN}$$

Where, TP=True Positive, TN = True Negative,
FP= False Positive, FN=False Negative.

For the Shenzhen dataset, different numbers of training and test images were utilized in the initial experiments, but experimentation showed that performance of the algorithm was almost similar if the training data was greater than or equal to 200. So, the number of training images were set to 200 while 40 images were used for validation and test sets. For JSRT dataset, 200 images are used for training, 20 are used for cross validation, and 20 images were used for testing. For the Montgomery dataset, 110 images for training, 10 images for cross validation and 18 images for testing were used. For all the experiments, the same number of epochs have been used, i.e., 350.

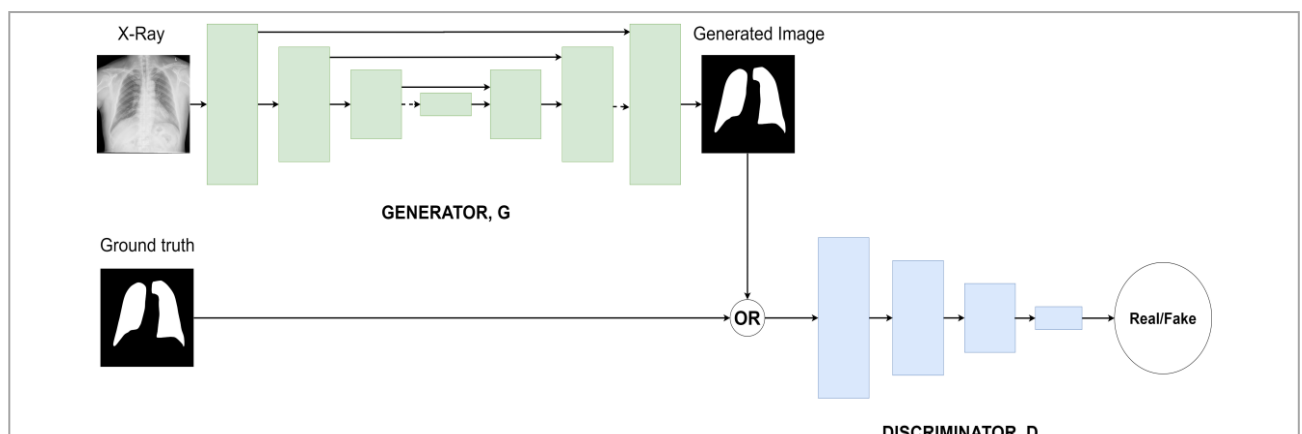

**FIGURE 4.** General Architecture of our GAN. 'G shows the architecture of our generator which is a U-net like network that maps a given X-ray image to its corresponding binary mask. Each block represents layers of network and an input image is down sampled until the middle layer after which up sampling starts. Skip connections are used which helps preserve the morphology and their detail is available in [16]. 'D' represents our discriminator network, which is a simple convolution neural network that takes as an input either the ground truth image or the image generated by the generator and performs binary classification, i.e. output will be either '0' or '1'.

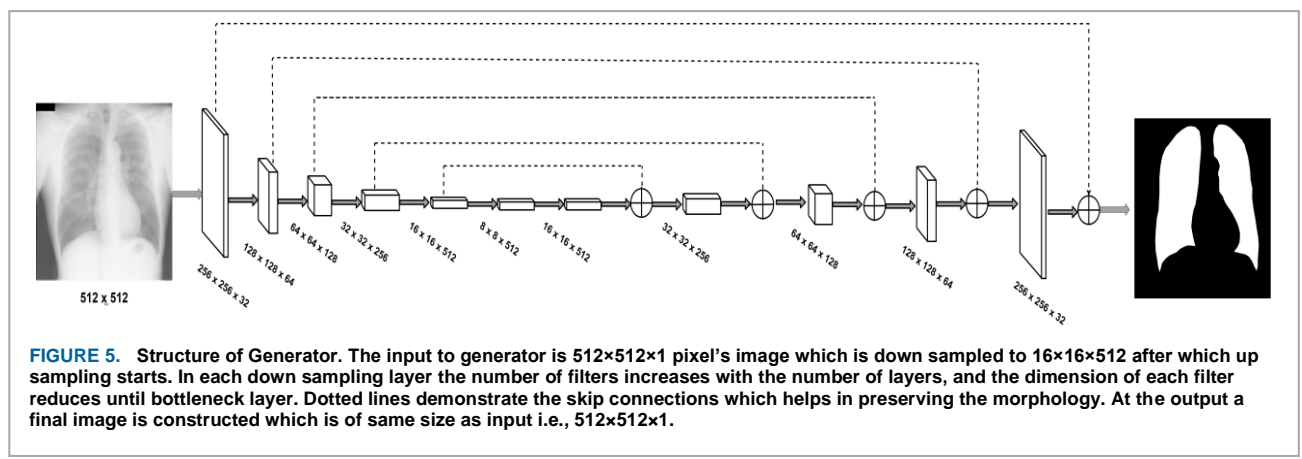

**FIGURE 5.** Structure of Generator. The input to generator is 512×512×1 pixel's image which is down sampled to 16×16×512 after which up sampling starts. In each down sampling layer the number of filters increases with the number of layers, and the dimension of each filter reduces until bottleneck layer. Dotted lines demonstrate the skip connections which helps in preserving the morphology. At the output a final image is constructed which is of same size as input i.e., 512×512×1.

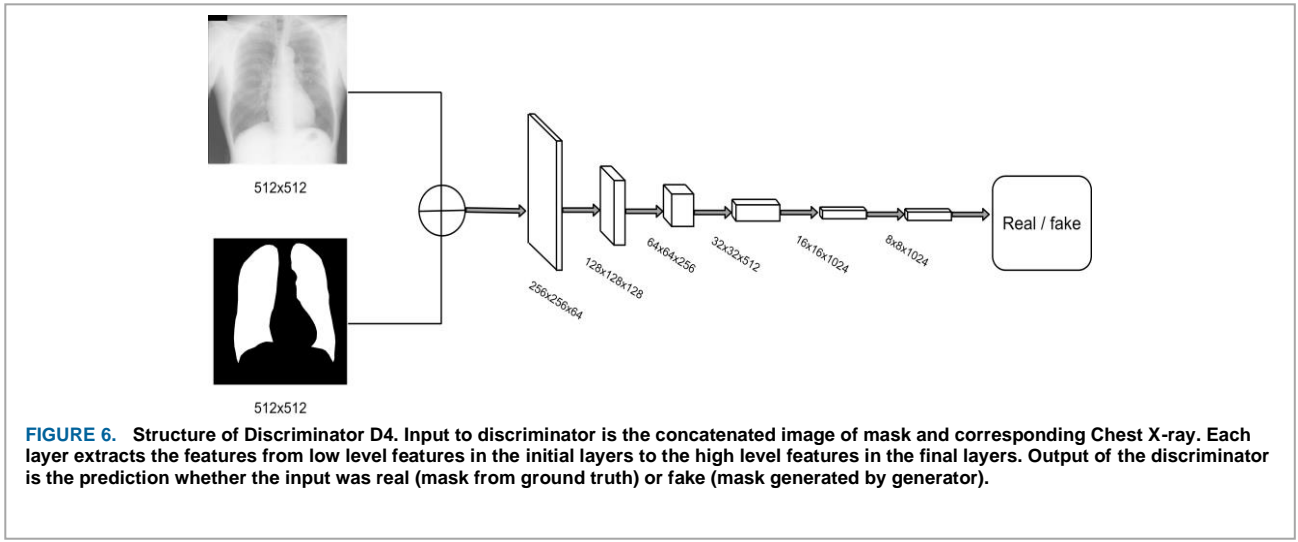

**FIGURE 6.** Structure of Discriminator D4. Input to discriminator is the concatenated image of mask and corresponding Chest X-ray. Each layer extracts the features from low level features in the initial layers to the high level features in the final layers. Output of the discriminator is the prediction whether the input was real (mask from ground truth) or fake (mask generated by generator).

**TABLE I**
**DATASET SPECIFICATIONS**

| Dataset Name | Number of images | | Image specifications |
|---|---|---|---|
| JSRT [45] | 154 normal | 93 TB | Image format: PNG<br>Image size: 2048 × 2048 pixels |
| Montgomery [46] | 80 normal | 58 TB | Image format: PNG<br>Image size: 4892 × 4020 |
| Shenzhen [46] | 326 normal | 336 TB | Image format: PNG<br>Size of the images varies for each X-ray.<br>The average size 3000 × 3000. |

**TABLE II**
**DICE SCORE COMPARISON OF D1, D2, D3 AND D4**

| Data set | Image Resolution | Dice Score | | | |
|---|---|---|---|---|---|
| | | D1 | D2 | D3 | D4 |
| Shenzen Chest X-ray Database | 256×256 | 0.9698 | 0.9694 | 0.9688 | 0.9698 |
| | 400×400 | 0.9667 | 0.9667 | 0.9671 | 0.9649 |
| | 512×400 | 0.9752 | 0.9747 | **0.9777** | 0.9737 |
| | 1024×1024 | 0.9750 | 0.9738 | 0.9770 | 0.9735 |
| Montgomery Chest X-Ray database | 256×256 | 0.9684 | **0.9780** | 0.9700 | 0.9730 |
| | 400×400 | 0.9696 | 0.9775 | 0.9673 | 0.9671 |
| | 512×400 | 0.9620 | 0.9628 | 0.9633 | 0.9671 |
| | 1024×1024 | 0.9617 | 0.9622 | 0.9630 | 0.9670 |

In Table 2, the performance of all the discriminators on the Shenzhen and Montgomery chest X-ray dataset are compared, while Figure 7 and 8 show sample segmented masks of our technique on both these datasets, respectively. The Montgomery dataset is very different from the Shenzen dataset in terms of the contrast of images and their bit depths. For a given dataset, the same number of training and test images for all the different configurations of the network along with all the sizes of images were used. The best result for each dataset is shown in bold. Overall, D3 has given the best segmentation result on the Shenzhen dataset, and D2 has given the best segmentation result on the Montgomery dataset.

Another observation is that with the increasing size of the images, the accuracy decreases slightly. The main reason for this is because as the size of the image increases, there are now more number of features to learn, and with the same number of layers, the accuracy is compromised.

For JSRT, our network is evaluated with discriminators D2 and D3, as they show good performance on the previous two datasets. Figure 9 shows sample segmented masks of our technique on the JSRT dataset. Subsequently, comparison is done between our results with other techniques that were implemented on the JSRT dataset. The same criteria were used for dividing the dataset into train, validation and test sets. For the sake of comparison with other techniques like [47], [48], [49], only the size of 400×400 pixels is used and the training is done over 350 epochs. The results in comparison with other reported techniques are reported in Table 3 and Table 4, which show our technique achieved the best results. Table 4 compares the IOU score with previous techniques. The discriminator 'D3' has achieved the best results by achieving 94.98 %

IOU score.

Figure 7 shows the test time comparison of our model for various image sizes used. The maximum time was taken by 1024×1024 pixels image which was 0.08 seconds. It is clear from Figure 7 that the required time for the test image increases if the test image size is increased, and the rise in time has correspondence with the rise in the image size.

TABLE III
DICE SCORE COMPARISON BETWEEN SAN, FCN, D2 AND D3 ON JSRT

| Technique Name | Dice Score |
|---|---|
| FCN [47] | 0.96 |
| SCAN [47] | 0.97 |
| Discriminator D2 (Ours) | 0.97 |
| **Discriminator D3 (Ours)** | **0.974** |

TABLE IV
IOU SCORE COMPARISON OF DIFFERENT TECHNIQUES ON JSRT

| Technique Name | Pre Processing | IOU (%) |
|---|---|---|
| Human Observer [50] | No | 94.6 |
| Registration [48] | No | 92.5 |
| ShRAC (Shape Regularized Active Contour) [49] | No | 90.7 |
| ASM (Active Shape Model) [50] | No | 90.3 |
| AAM (Active Appearance Model) [50] | No | 84.7 |
| Mean Shape [50] | No | 71.3 |
| FCN (Fully Convolutional Neural Network) [47] | No | 92.9 |
| SCAN (Structure Correcting Adversarial Network) [47] | No | 94.7 |
| **D3 (Ours)** | **No** | **95** |

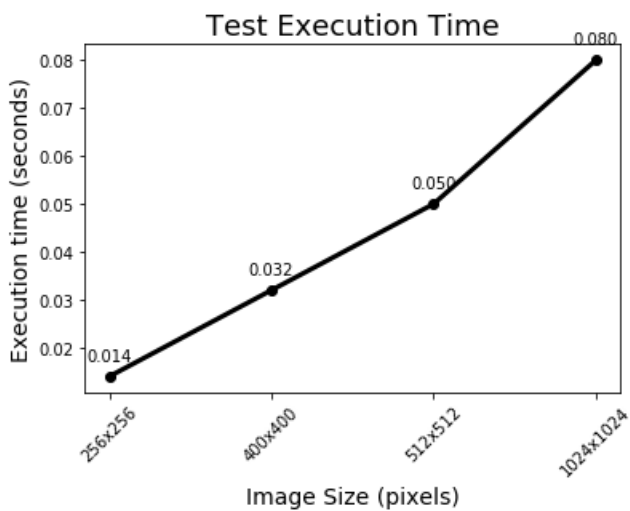

FIGURE 7. Test time comparison between different sizes of test images and their execution time. Increasing the test image size correspondingly increase the test time.

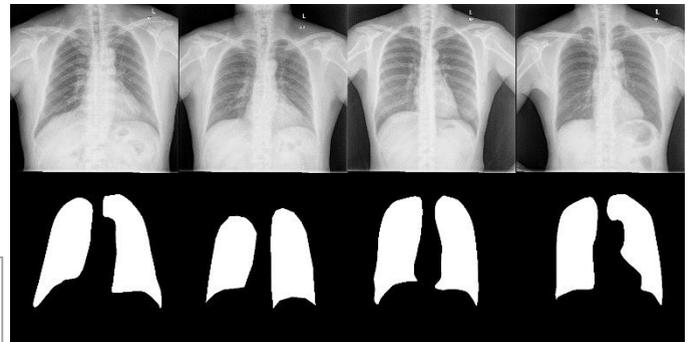

FIGURE 8. X-ray images and segmented masks of Shenzhen dataset.

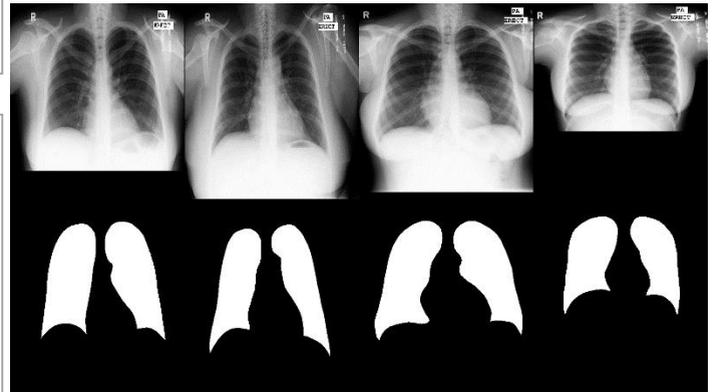

FIGURE 9. X-ray images and segmented masks of JSRT X-ray images and segmented masks of Montgomery dataset.

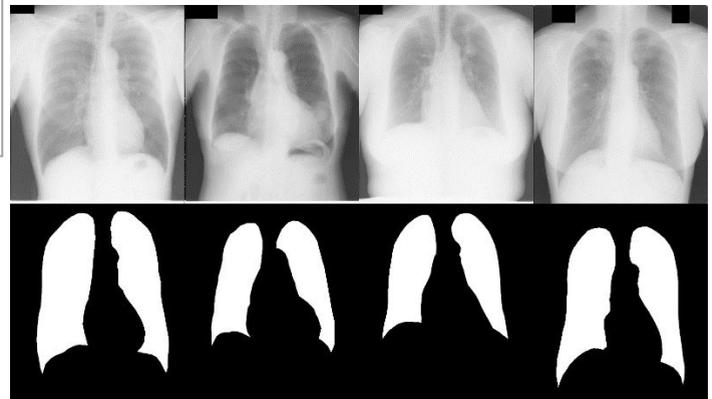

FIGURE 10. X-ray images and segmented masks of JSRT dataset.

### A. OVER-SEGMENTATION AND UNDER-SEGMENTATION

Generally, there are two cases of anomalies in segmentation tasks; Over-segmentation, and Under-segmentation. Over-segmentation occurs if pixels of non-target objects are classified as the objects of interest. For lung segmentation, the over-segmentation is the scenario when some or all the pixels of the non-lung region (object of non-interest) are classified as the lungs' region (object of interest). Under-segmentation refers to the scenario when some or all the pixels of the lungs' region are classified as background pixels. Typically, in the case of over-segmentation, it might still be possible to reconstruct the lungs' region because no useful data is lost. However, in the case of under-segmentation, key lungs' region can be lost. Hence, the

VOLUME XX, 2020

under-segmentation must be avoided [55]. There can be multiple reasons for over and under segmentation. For example, the same color or gradient of the object and the background make it challenging to generate a robust segmentation boundary.

In the case of lung CXR's, the color or gradient of lungs is very similar to that of the non-human background, so the task of segmenting is considered very challenging. Figure 11 shows the cases of over and under segmentation. As can be seen in input images, the area surrounding the lungs has a smooth intensity variation, and thus, results in over-segmentation in (a), (b) and (c). A slight under-segmentation is observed in (d) because of very low contrast in the lower lungs' region.

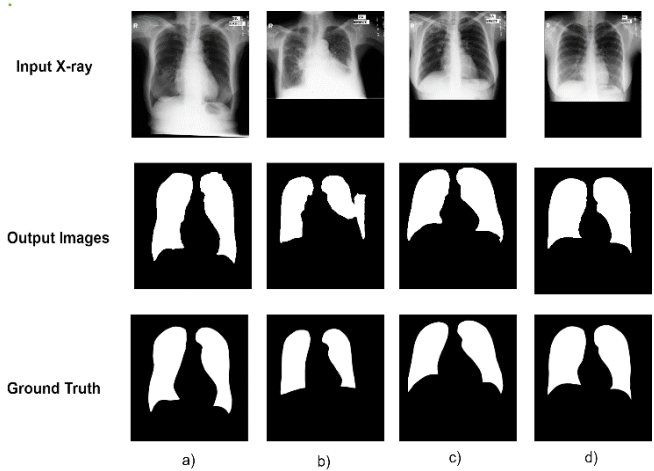

FIGURE 11. Example cases for over-segmentation and under-segmentation. First row shows the input images to the generator, second row shows the output images generated by the generator and third row shows their respected ground truth masks. (a), (b), (c) show over-segmentation. (d) shows under-segmentation attributed to the very low contrast in the lower lung region.

## V CONCLUSION

As computer aided diagnostic systems (CADs) are getting popular among doctors and health care professionals, it is fundamentally important to have a reliable segmentation technique for CXRs so that the images are interpreted correctly and can be used for potential diagnosis applications. In this study, lung image segmentation is performed from chest x-ray images using generative adversarial networks. Results are reported for four different types of discriminators, and a comparison of the results with other state-of-the-art techniques is presented. Experimental results are reported on three different datasets, namely; JSRT (247 images), Montgomery dataset (138 images) and Shenzhen dataset (566 images). Among the four discriminators, the discriminator D3, i.e., 70×70 patch-GAN has outperformed other discriminator networks in terms of dice score. Our method in this work has achieved state-of-the-art results for lung segmentation compared to other methods. Comparisons with other state-of-the-art methods also show improvement for the discriminator D3 in terms of IOU score. A limitation of the proposed generative adversarial network is that it is resource expensive and requires high computing power. Besides, the proposed model has been trained and evaluated for three publicly available CXR datasets. For new datasets, variations in the image acquisition and imaging standards are not unexpected, as different manufacturers may have slightly different image outcome. For such newer datasets, a new study would be required, and it is yet to be determined how well the proposed model generalizes for newer datasets – similar to many deep learning techniques.